\documentclass[aps,11pt,twoside]{revtex4}

\usepackage{amsmath,latexsym,amssymb,verbatim,enumerate,graphicx}

\usepackage{float}
\usepackage[english]{babel}
\usepackage[T1]{fontenc}
\usepackage{subfigure}
\usepackage{float}
\usepackage{appendix}
\usepackage{pdfpages}
\usepackage{tikz}
\usepackage{braket}
\usepackage{anyfontsize}
\usepackage[11pt]{moresize}
\usepackage{enumerate}
\usepackage{dcolumn}
\usepackage{bm}
\usepackage{tabularx}
\usepackage{array}
\usepackage{lipsum}
\usepackage{cancel}
\usepackage{enumitem}

\usepackage{theorem}

\usetikzlibrary{quantikz}

\newcommand{\cN}{{\cal N}}
\newcommand{\be}{\begin{equation}}
\newcommand{\ee}{\end{equation}}
\newcommand{\bea}{\begin{eqnarray}}
\newcommand{\eea}{\end{eqnarray}}

\begin{document}

\title{Discriminating qubit amplitude damping channels}

\author{Milajiguli Rexiti}
\email{milajiguli.milajiguli@unicam.it}
\affiliation{School of Science and Technology, University of Camerino, 62032 Camerino, Italy}

\author{Stefano Mancini}
\email{stefano.mancini@unicam.it}
\affiliation{School of Science and Technology, University of Camerino, 62032 Camerino, Italy \\
and INFN-Sezione di Perugia, 06123 Perugia, Italy}

\date{\today}

\begin{abstract}
We address the issue of the discrimination between two-qubit 
amplitude damping channels by exploring several strategies.
For the single-shot, we show that the excited state does not always give the optimal input, and that side entanglement assistance has limited benefit.
On the contrary, feedback assistance from the environment is more beneficial.
For the two-shot, we prove the in-utility of entangled inputs. 
Then focusing on individual (local) measurements, we find the optimal adaptive strategy.
\end{abstract}

\maketitle

\section{Introduction}

A quantum channel is a linear stochastic (precisely a linear, completely positive and trace-preserving) map on the set of density operators \cite{bookMW}. As such, it can describe any physical process. It is often essential to distinguish between two (or even more) physical processes. Hence the issue of quantum channel discrimination becomes pervasive well beyond the boundary of information theory \cite{Acin,GLN,MFS,WY,Hay}.
Quite generally, channel discrimination is a challenging task \cite{CPR,Wat,Sacchi,max}. In fact, although it can be traced back to the (somehow old problem of) states discrimination \cite{Hel,Hol}, 
it involves a double optimization: on the output measurement and the input state.
Recently, bounds on the error probability were found for general strategies \cite{Pirs,W20}.

Among quantum channels, the amplitude damping plays a prominent role as it describes the
energy loss of a system, which is the most common effect occurring in an open system.
As a matter of fact, this channel is often invoked as an example when dealing with discrimination (see e.g. \cite{Pirs,W20}).
However, a systematic and thorough study of amplitude damping channel discrimination starting from dimension two is still lacking. Here we address this issue and unveil several unexpected results. For one-shot discrimination, the optimal input state turns out to not always be the excited state.
Furthermore, side entanglement has a limited benefit because of a limited parameter region where it brings improvement and the smallness of such improvement. In contrast, feedback assistance from the  environment results more beneficial. By such feedback, 
we mean the possibility to access the environment,
measure it and then use this (classical) information to adjust the state of (or the measurement
process in) the main system according to the desired goal \cite{fbmodel}.
Additionally, we prove that entangled inputs are not useful for two-shot, although collective measurement can give the minimum error probability. Then, restricting the attention to individual measurements, we find the optimal adaptive strategy (useful for a complete LOCC strategy \cite{Tej}).

The paper is organized in two main Sections.
Sec.\ref{sec:oneshot} is devoted to one-shot discrimination.
In Subsec.\ref{subsec:opt1} the optimal input is found.
In Subsec.\ref{subsec:sideent} side entanglement is considered.
Feedback assistance model is presented in Subsec.\ref{subsec:fb}.
Then, Sec.\ref{sec:twoshot}  is devoted to two-shot discrimination.
In Subsec.\ref{subsec:opt2} the optimal input is found when also collective measurements are allowed.
The optimal adaptive strategy for individual measurements is devised in Subsec.\ref{subsec:ada}.
Finally, in Sec.\ref{sec:con} conclusions are drawn.


\section{One-shot discrimination}\label{sec:oneshot}

It is customary to consider the amplitude damping channel as coming from the unitary interaction of the system with the environment generated by
\be
H=\eta\left(a^\dag b+a b^\dag\right),
\ee
where $a,a^\dag$ (res. $b,b^\dag$) are the ladder operators of the system (resp. environment).

For a qubit system and qubit environment, in the computational basis $\{|00\rangle, |01\rangle,
|10\rangle, |11\rangle\}$ we have 
\be
H=\begin{pmatrix}
0 & 0 & 0 & 0 \\
0 & 0 & \eta & 0 \\
0 & \eta & 0 & 0 \\
0 & 0 & 0 & 0
\end{pmatrix},
\ee
and then the corresponding unitary $U=e^{-iH}$ reads
\be\label{Uqubit}
U=\begin{pmatrix}
1 & 0 & 0 & 0 \\
0 & \cos\eta & -i\sin\eta & 0 \\
0 & -i\sin\eta & \cos\eta & 0 \\
0 & 0 & 0 & 1
\end{pmatrix}.
\ee
The map on the system's states can be written as
\be\label{calN}
\cN(\rho)=K_0\rho K_0^{\dag}+K_1\rho K_1^{\dag},
\ee
where the Kraus operators are given by
\begin{align}
K_0&=\langle 0| U |0\rangle =
{ \begin{pmatrix} 1 & 0\\ 0 & \cos{\eta}
\end{pmatrix},}
\\
K_1&=\langle 1| U |0\rangle = 
{
\begin{pmatrix} 0 & -i\sin{\eta}\\ 0 & 0
\end{pmatrix},} 
\end{align}
with the bra-ket taken on the environment. The quantity $\sin^2\eta$  represents the decay probability ($\eta\in[0,\frac{\pi}{2}]$).


\subsection{Optimal input}\label{subsec:opt1}

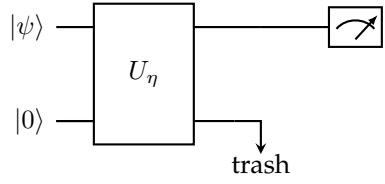
\begin{figure}[H]
\centering
\begin{quantikz}
\lstick{$\ket{\psi}$}& \gate[wires=2]{U_{\eta}}{2cm}& \qw & \meter{}  \\ 
\lstick{$\ket{0}$}&\qw& \trash{\text{trash}} 
\end{quantikz}    
      \caption{Schematic representation of the channel discrimination through unitary dilation. One has to determine whether $U_{\eta_0}$ or $U_{\eta_1}$ acted by controlling only the main system (top line).}
      \label{fig1}
\end{figure}

Let us analyze the distinguishability of two amplitude damping channels
characterized by parameters $\eta_0$ and $\eta_1$. By referring to Fig.\ref{fig1}, we 
assume that each one acts with probability $P_{0}=P_{1}=1/2$ on an input state
\begin{equation}\label{ketpsi}
\ket\psi=\sqrt{1-x}\ket{0}+e^{-i\varphi}\sqrt{x}\ket{1},
\end{equation}
where $ x\in[0,1]$ 
 and $ \varphi\in[0, 2\pi)$.
Without loss of generality, we take $\eta_0>\eta_1$.
We can also set $\varphi=0$, because of the symmetric action of ${\cal N}$ 
with respect to the $z$ axis in the Bloch sphere.

\begin{figure}[H]
\centering
\includegraphics[width=6cm]{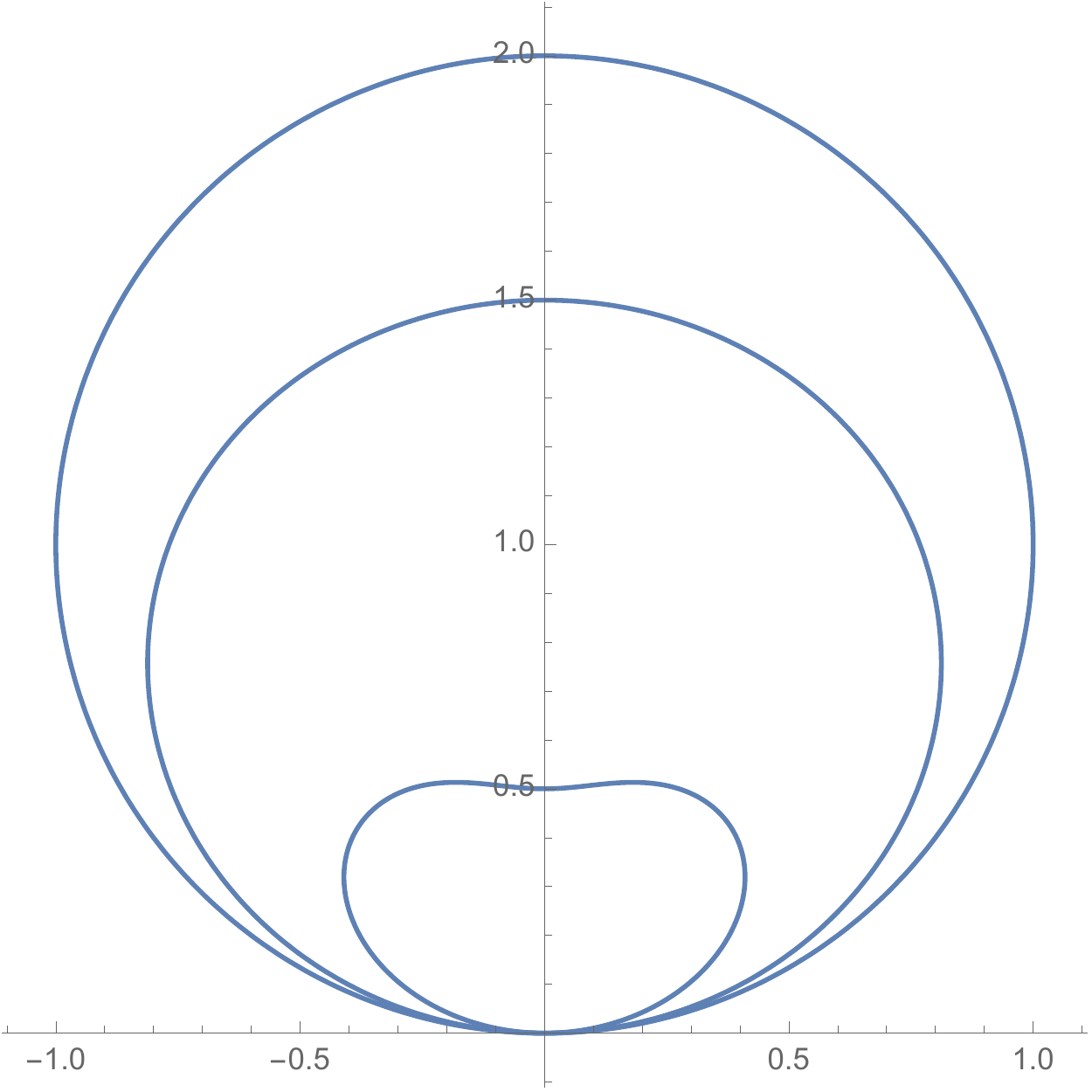}
\caption{Pictorial representation of the amplitude damping effect. Points along the curves have polar coordinates $\| \ket 0\bra 0-\rho_1(\theta)\|_1$ and $\theta$, where $\theta\equiv\arcsin \sqrt x$. Different curves corresponds to different values of $\eta_1$ (from outer to inner curve $\eta_1=0$, $\pi/6$ and $\pi/3$). The point at the intersection of the vertical axes and each curve above the origin corresponds to the state 
$\rho_1={\cal N}(|1\rangle\langle 1|)$.}
\label{fig2new}
\end{figure}

According to \eqref{Uqubit}, we obtain
\bea
U_{\eta_i}\ket\psi\ket{0}&=& \sqrt{1-x}\ket{00}+\sqrt{x}\left(-i\sin{\eta_i}\ket{01}+\cos{\eta_i}\ket{10}\right),
\qquad i=0,1.
\label{U2}
\eea
Consequently, the output state of the channel reads
\begin{align}
\rho_{i}&=\left(1-x +x \sin^2\eta_i \right)\ket0 \bra0+\sqrt{x(1-x)}\cos\eta_i\ket0 \bra1\notag\\
&+\sqrt{x(1-x)}\cos\eta_i\ket1 \bra0+x\cos^2\eta_i \ket1 \bra1, 
\qquad i=0,1.
\label{rho2}
\end{align}
So the problem of channel discrimination is now translated into the discrimination between two mixed states, $\rho_{0}$ and $\rho_{1}$, each occurring with probability $1/2$.
To this end, it is known that the optimal measurement is given by the {projection on the 
positive and negative subspaces of 
$\rho_0-\rho_1$ \cite{bookMW} (also known as Helstrom measurement \cite{Hel}).} Denoting by $|v_0\rangle$ and $|v_1\rangle$ its (normalized) eigenvectors corresponding respectively to positive and negative eigenvalues, 
{we can evaluate
\begin{subequations}\label{contributionsPs}
\begin{align}
P_0(0)&\equiv \langle v_0|\rho_0|v_0\rangle,\\
P_1(1)&\equiv \langle v_1|\rho_1|v_1\rangle.
\end{align}
\end{subequations}
Then, the probability of success in discriminating between 
$\rho_{0}$ and $\rho_{1}$, each occurring with probability $1/2$, is given by 
\be\label{Psuccess}
P_{succ}=\frac{1}{2}P_0(0)+\frac{1}{2}P_1(1).
\ee
This turns also out to be \cite{bookMW}
\be\label{sucpr}
P_{succ}=\frac{1}{2}\left(1+\frac{1}{2}\left\| \rho_{0}- \rho_{1} \right\|_1\right),
\ee
where $\| T \|_1\equiv {\rm Tr}\sqrt{T^{\dag}T}$.

Inserting \eqref{contributionsPs} into \eqref{Psuccess} (or equivalently using \eqref{rho2} into \eqref{sucpr}), yields explicitly
\begin{align}\label{Psuccx}
P_{succ}=\frac{1}{2}\left\{1+(\cos\eta_1-\cos\eta_0)\sqrt{x\left[1-x(1-\gamma^2)\right]}\right\},
\end{align}
where
\begin{align}
\gamma\equiv\gamma(\eta_0,\eta_1)\equiv \cos\eta_1+\cos\eta_0.
\label{gamma} 
\end{align}
Maximizing \eqref{Psuccx} over $x$ (taking into account that $0\leq x\leq 1$) we get 
}
\begin{subequations}\label{Psuccnofb}
\begin{align}
P_{succ}&=\frac{1}{4}\left(2+ \frac{ \cos \eta_1-\cos\eta_0}{\sqrt{1-\gamma^2}}\right),
\quad \gamma < \frac{1}{\sqrt 2}, \\
P_{succ}&=\frac{1}{2}\left( \sin^2\eta_0+\cos^2\eta_1\right),
\quad \gamma \geq \frac{1}{\sqrt 2}.
\end{align}
\end{subequations}
Note that the optimal value of $x$ results 
\begin{subequations}\label{optx1shot}
\begin{align}
x&=\frac{1}{2(1-\gamma^2)},
\quad \text{for} \quad \gamma<\frac{1}{\sqrt{2}}, \\
x&=1,
\quad \text{for} \quad \gamma\geq\frac{1}{\sqrt{2}}.
\end{align}
\end{subequations}
This means, interestingly, that the optimal input state is not always the excited state $|1\rangle$ as one would expect (just because it is commonly considered as the most sensitive state to the damping action).
{The result can be better understood by looking at Fig.\ref{fig2new}.  {There we show the separation between $\rho_0$, for $\eta_0=0$ (i.e., $\rho_0=\ket 0 \bra 0$) and $\rho_1$ in Eq.\eqref{rho2}, for different values of $\eta_1$.}
One can clearly see that once $\eta_1$ becomes bigger than $\pi/4$
(hence $\gamma$ becomes smaller than $\frac{1}{\sqrt{2}}$), the farthest point from the origin is no longer on the {vertical} axes (see curve corresponding to $\eta=\pi/3$).}

\subsection{Side entanglement}\label{subsec:sideent}

\begin{figure}[ht]
\centering
\begin{quantikz}
\makeebit[-60]{$\ket{\psi}$} &\qw&\qw&\qw\rstick[wires=2]{Measurement}\\
& \gate[wires=2]{U_{\eta}}{2cm}& \qw & \qw \\ 
\lstick{$\ket{0}$}& \qw & \trash{\text{trash}}      & & 
\end{quantikz}    
      \caption{Model for channel discrimination exploiting entanglement between the input system and an accessible reference system. }
      \label{fig2}
\end{figure}
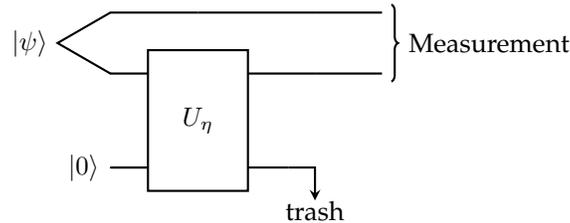

Consider the usage of side entanglement according to the model of Fig.\ref{fig2}.

Let $\ket{\Psi}=\sqrt{1-{y}}\ket{01}+\sqrt{y}\ket{10}$ be an entangled state between the reference system and the channel's input. 
Then, we have to distinguish between the following two states at the measurement stage:
\begin{align}\label{rhooutside}
{\rm Tr}_E\left\{ \left(I\otimes U_{\eta_i} \right)\ket\Psi \ket 0 \bra 0 \bra\Psi
 \left(I\otimes U_{\eta_i} \right)^\dag \right\} 
 &=\cos^2\eta_i (1-{y}) |01\rangle\langle 01|+{y} |10\rangle\langle 10| \notag\\
& +\sqrt{{y}(1-{y})} \cos\eta_i
 \left(|01\rangle\langle 10|+|10\rangle\langle 01|\right)\notag\\
 &+(1-{y})\sin^2\eta_i |00\rangle\langle 00|,
 \qquad i=0,1,
 \end{align}
 where the unitaries are as in \eqref{Uqubit}.
  
\begin{figure}[H]
\centering
\includegraphics[width=9cm]{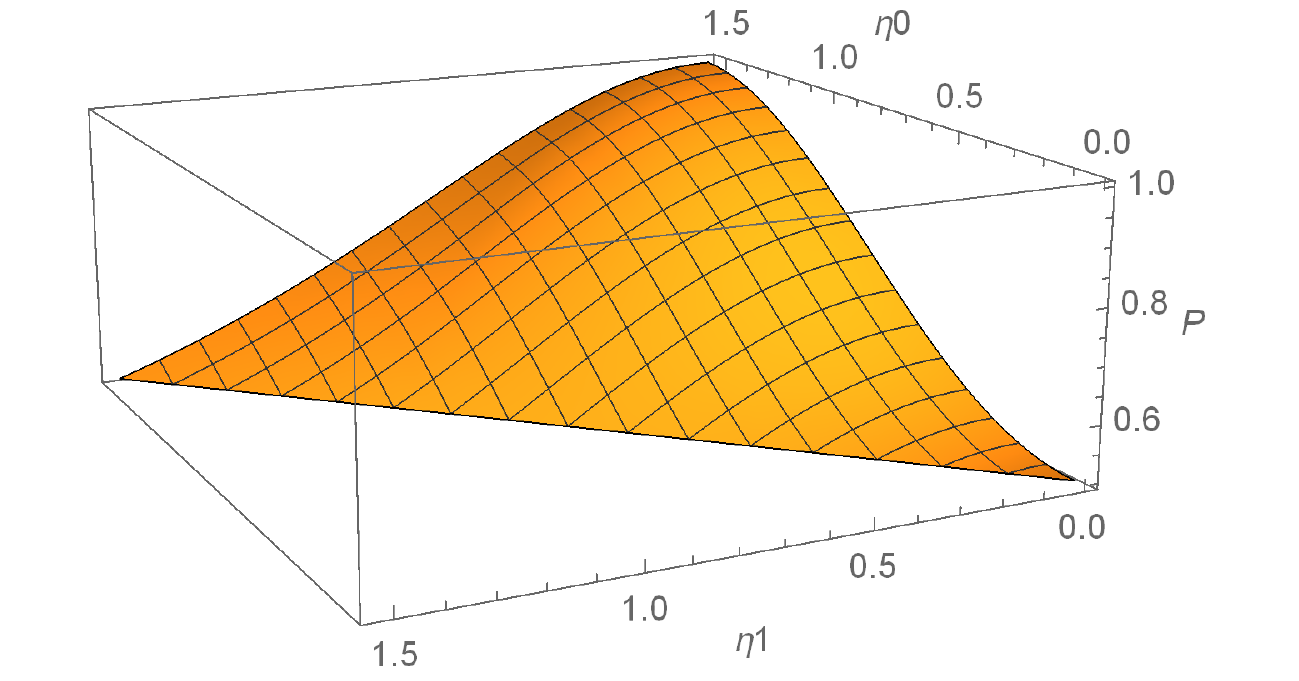}
\caption{Maximum probability of success $P(\eta_0,\eta_1,y^*)$ vs $\eta_0$ and $\eta_1$.}
\label{fig4new}
\end{figure}

\begin{figure}[H]
\centering
\includegraphics[width=8cm]{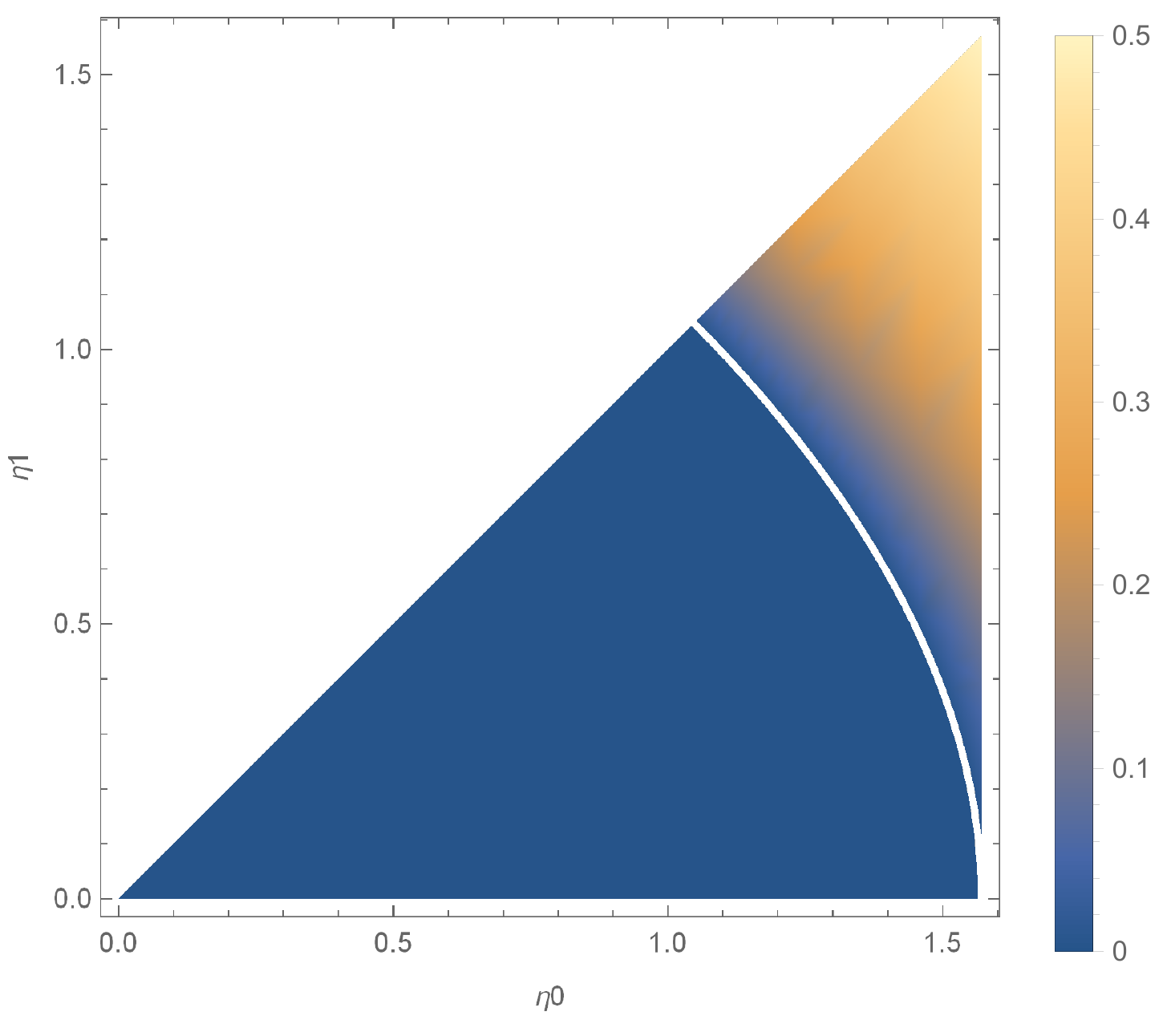}
\caption{Optimal value ${y}^*$ (Eq.\eqref{eq:xoptDelta}) vs $\eta_0$ and $\eta_1$. {The white line corresponds to $\frac{\gamma-1}{\gamma-2}=0$}.}
\label{fig4}
\end{figure}

The success probability \eqref{sucpr} for the states in \eqref{rhooutside} leads to 
{
\begin{align}\label{eq:Deltanofb}
P_{succ}(\eta_0,\eta_1,y)=\left(\cos\eta_1-\cos\eta_0\right)\left\{(1-y)\gamma+\sqrt{(1-y)\left[4y+(1-y)\gamma^2\right]} \right\}.
\end{align}
Then, in Fig.\ref{fig4new} we show $P_{succ}(\eta_0,\eta_1,{y}^*)$,
}
where 
\begin{equation}\label{eq:xoptDelta}
{y}^*(\eta_0,\eta_1):={\rm argmax}_{y} \; P_{succ}(\eta_0,\eta_1,{y})
{= \max\left\{0,\frac{\gamma-1}{\gamma-2}\right\}.}
\end{equation}
This latter quantity gives the optimal amount of entanglement and is shown in Fig.\ref{fig4}.


\begin{figure}[H]
\centering
\includegraphics[width=8cm]{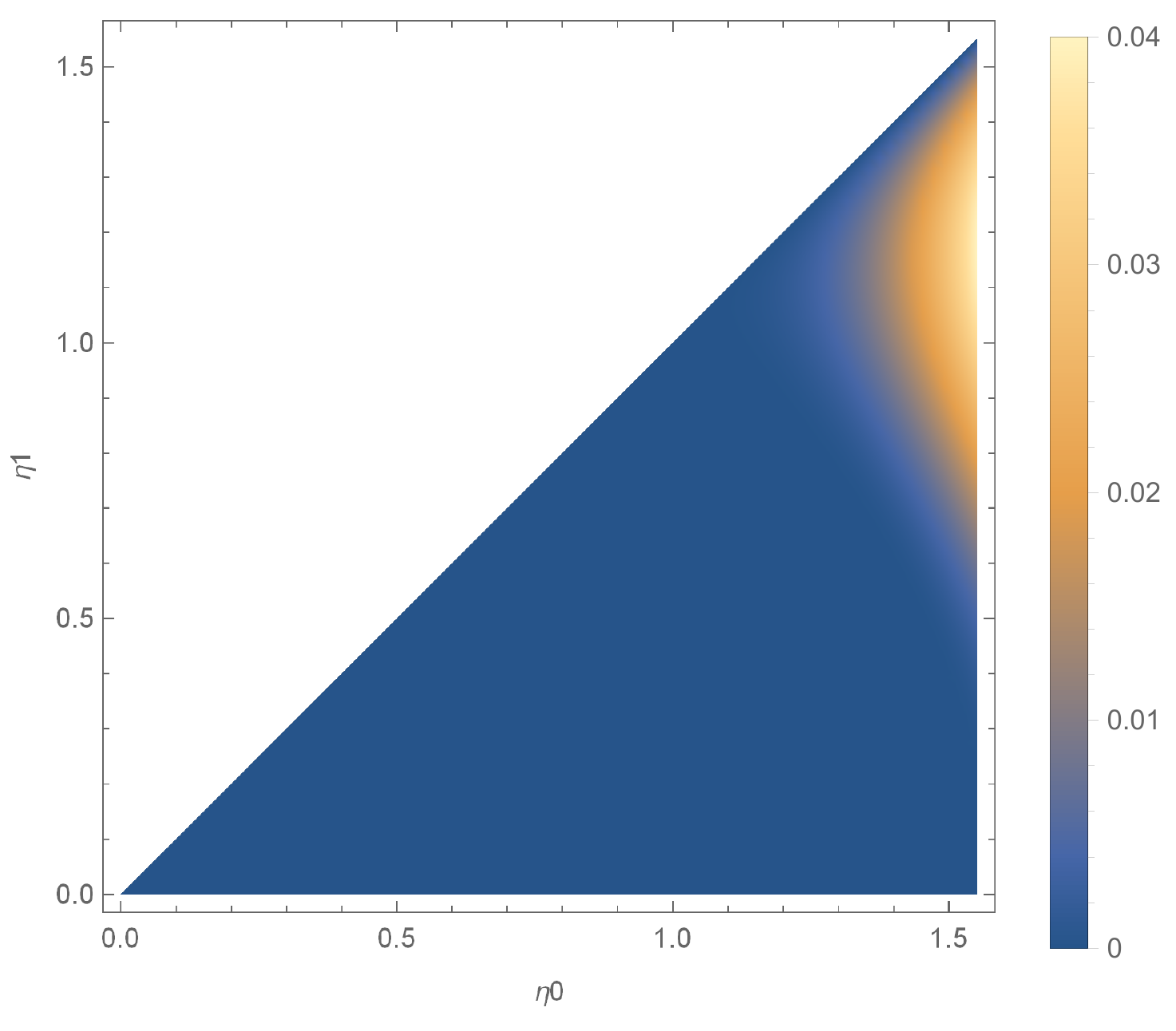}
\caption{Difference between success probability computed at optimal ${y}$ and ${y}=0$ (Eq.\eqref{eq:Deltanofb}) vs $\eta_0$ and $\eta_1$.}
\label{fig3}
\end{figure}

{
In Fig.\ref{fig3} we report the difference 
$P_{succ}(\eta_0,\eta_1,{y}^*)-P_{succ}(\eta_0,\eta_1,0).$}
It shows that the improvement due to the side entanglement is relatively tiny. Moreover, it does not occur in all parameters' region. 
{It is worth reminding that in other channels, like depolarizing channels, the effectiveness of side entanglement shows up for values of parameters making the channels entanglement-breaking \cite{Sacchi}. 
Here, we have amplitude damping channel that is never entanglement-breaking, 
except for $\eta_i=\pi/2$.
Nevertheless, the region with nonzero values in Fig.\ref{fig3} is close to this border.
It is also interesting to note from Fig.\ref{fig4} that a small amount of entanglement is more effective  than the maximal ($y$ never reaches the value $1/2$).

Finally, we note that according to the discussion at the end of SubSec.\ref{subsec:opt1}, the component 
$|10\rangle$ of $|\Psi\rangle$ (inputting no excitation to the channel) only plays a role when approaching the region $\gamma<1/\sqrt{2}$.
Then, the more we move towards the corner where $\eta_0=\eta_1=\pi/2$, the more the weight of the two components 
of $|\Psi\rangle$ become balanced. 
}

\subsection{The use of feedback}\label{subsec:fb}

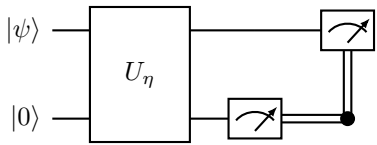
\begin{figure}[ht]
\centering
\begin{quantikz}
\lstick{$\ket{\psi}$}& \gate[wires=2]{U_{\eta}}{2cm}& \qw& \meter{} \\ 
\lstick{$\ket{0}$}   &  & \meter{}         &  \cwbend{-1}
\end{quantikz}    
      \caption{Channel discrimination with feedback. A first measurement is performed on the environment (bottom line), and then according to its outcome, a second measurement is performed on the system (top line). Here and in the following figures, double lines represent classical information.}
      \label{fig5}
\end{figure}

Let us now move on to a discrimination strategy using feedback as illustrated in Fig.\ref{fig5}.
It presumes the possibility to locally access the environment \cite{fbmodel}.

We perform on the environment a measurement in the basis
$\{\ket {\alpha^{+}},\ket {\alpha^{-}}\}$, where
{
\begin{subequations}
\begin{align}
\ket {\alpha^{+}}&=\cos \alpha \ket 0+\sin \alpha \ket 1, \\
\ket {\alpha^{-}}&=-\sin \alpha \ket 0+\cos \alpha \ket 1,
\end{align}
\end{subequations}
}
are two orthogonal states in the plane $x-z$ with $\alpha\in[0,\pi/2]$ to be determined. 

Then, depending on the environment's measurement result, we choose a proper observable for the system. In this way, the feedback \emph{actuation} has to be intended as the measurement on the main system performed conditioned to the environment's measurement outcome.

\begin{figure}[H]
\centering
\includegraphics[width=8cm]{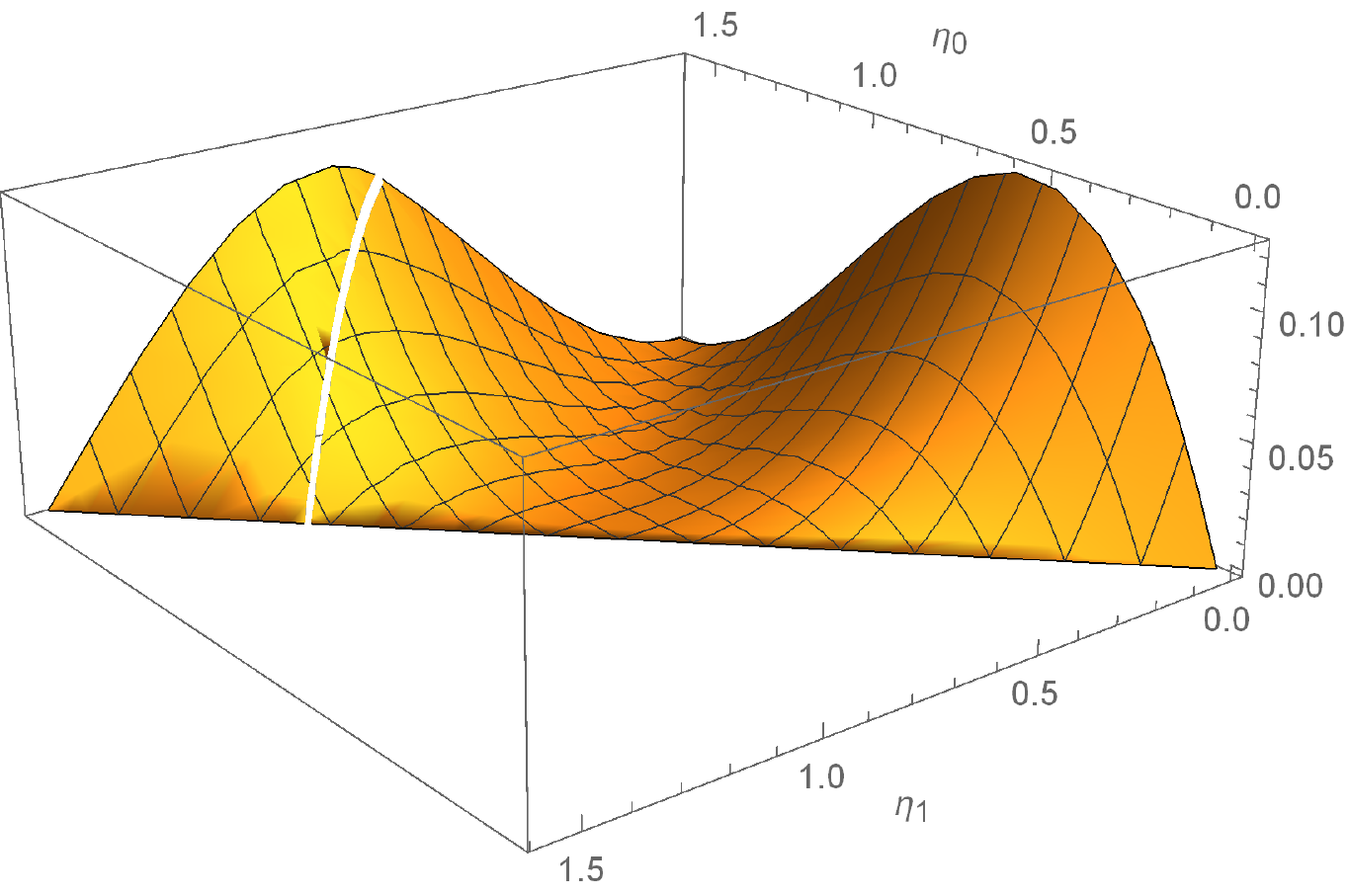}
\caption{Difference between the maximum probability of success with feedback \eqref{Psuccfb} and without feedback \eqref{Psuccnofb} vs $\eta_0$ and $\eta_1$. The white line corresponds to $\gamma=1/\sqrt{2}$ (see \eqref{gamma}).}
\label{fig6}
\end{figure}

We can hence distinguish between the following possibilities:
\begin{itemize}
\item Environment outcome +1 (eigenvalue corresponding to $\ket{\alpha^{+}}$). This happens with probability:
\begin{align}
P(+1)&=\frac{1}{2}{\rm Tr}\left(\bra{\alpha^{+}} U_{\eta_0}\ket{\psi}\ket{0}\bra{0}\bra{\psi} 
U_{\eta_0}^\dag \ket{\alpha^{+}}\right) \notag\\
&+\frac{1}{2}{\rm Tr}\left(\bra{\alpha^{+}} U_{\eta_1}\ket{\psi}\ket{0}\bra{0}\bra{\psi} 
U_{\eta_1}^\dag \ket{\alpha^{+}}\right),
\end{align}
and the resulting state on the system is
\be
\ket{\varphi_0^{+}}=\frac{1}{N_0^{+}} \bra {\alpha^{+}} \Big(U_{\eta_0}\ket\psi\ket0 \Big),
\label{varphi1p} 
\ee
or 
\be
\ket{\varphi_1^{+}}=\frac{1}{N_1^{+}} \bra {\alpha^{+}} \Big(U_{\eta_1}\ket\psi\ket0 \Big),
\label{varphi2p}
\ee
depending on which unitary has acted.
Here, $1/ N_i^{+}, i=0,1 $ are normalization factors. 
It is then optimal to measure the observable $\ket{\varphi_0^{+}}\bra{\varphi_0^{+}}-\ket{\varphi_1^{+}}\bra{\varphi_1^{+}}$ on the main system to discriminate between \eqref{varphi1p} and \eqref{varphi2p} \cite{bookMW}. This can be done, following \eqref{sucpr}, with probability\footnote{This expression derives from \eqref{sucpr} when employing pure state vectors.} $\frac{1}{2}\left( 1+\sqrt{1-| \braket{\varphi_0^{+}|\varphi_1^{+}}|^2}\right) $.
Thus, the probability of success when environment outcome is $+1$, reads:
\be
P_{succ}^{(+)}=P(+1)\times\frac{1}{2}\left( 1+\sqrt{1-| \braket{\varphi_0^{+}|\varphi_1^{+}}|^2}\right). 
\label{Psp}
\ee
\item Environment outcome -1 (eigenvalue corresponds to $\ket{\alpha^-}$).
 This happens with probability:
\begin{align}
P(-1)&=\frac{1}{2}{\rm Tr}\left(\bra{\alpha^{-}} U_{\eta_0}\ket{\psi}\ket{0}\bra{0}\bra{\psi} 
U_{\eta_0}^\dag \ket{\alpha^{-}}\right) \notag\\
&+\frac{1}{2}{\rm Tr}\left(\bra{\alpha^{-}} U_{\eta_1}\ket{\psi}\ket{0}\bra{0}\bra{\psi} 
U_{\eta_1}^\dag \ket{\alpha^{-}}\right),
\end{align}
and the resulting state on the system is
\be
\ket{\varphi_0^{-}} =\frac{1}{N_0^{-}} \bra {\alpha^{-}} \Big(U_{\eta_0}\ket\psi\ket0 \Big),
\label{varphi1m}
\ee
or
\be
\ket{\varphi_1^{-}}=\frac{1}{N_1^{-}} \bra {\alpha^{-}} \Big( U_{\eta_1}\ket\psi\ket0 \Big),
\label{varphi2m}
\ee
depending on the acted unitary.
Here $ 1/ N_i^{-}, i=0,1 $ are normalization factors. It is then optimal to measure the observable $\ket{\varphi_0^{-}}\bra{\varphi_0^{-}}-\ket{\varphi_1^{-}}\bra{\varphi_1^{-}}$ on the main system to discriminate between \eqref{varphi1m} and \eqref{varphi2m}. 
Thus, the probability of success when environment outcome is $-1$, reads:
\be
P_{succ}^{(-)}=P(-1) \times \frac{1}{2}\left( 1+\sqrt{1-| \braket{\varphi_0^{-}|\varphi_1^{-}}|^2}\right). 
\label{Psm}
\ee
\end{itemize}
Finally, putting \eqref{Psp} and \eqref{Psm} together we get the overall probability of success as
\bea
P_{succ}&=&P_{succ}^{(+)}+P_{succ}^{(-)}\notag\\
&=&\frac{\chi}{2}\left[ 1+\sin \alpha \sin \left(\frac{\eta_0-\eta_1}{2} \right) \sqrt{\frac{x\left(\mu +\nu\right)}{c_1c_2}}\right]\notag\\
&+&\frac{1-\chi}{2}\left[ 1+\cos \alpha \sin \left(\frac{\eta_0-\eta_1}{2} \right) \sqrt{\frac{x\left(\mu -\nu\right)}{(1-c_1)(1-c_2)}}\right],
\label{Pfs}
\eea
where
\begin{subequations}
\begin{align}
\chi&\equiv
\frac{1}{2}-\frac{1}{2}\cos(2\alpha)\left[1-x\left(\sin^2\eta_0+\sin^2\eta_1\right)\right],
\\
\mu&\equiv\left[1+(2x-1)\cos\eta_0\cos\eta_1+\sin\eta_0\sin\eta_1\right],
\\
\nu&\equiv\cos(2\alpha)\left[(2x-1)+\cos\eta_0\cos\eta_1+(2x-1)\sin\eta_0\sin\eta_1\right],
\\
c_i&\equiv\frac{1}{2}-\frac{1}{2}\cos(2\alpha)\left[1-2x\sin^2\eta_i\right],
\qquad i=0,1.
\end{align}
\end{subequations}
Analyzing Eq.\eqref{Pfs}, we found the maximum success probability as 
\be\label{Psuccfb}
P_{succ}=\frac{1+\sin (\eta_0-\eta_1)}{2},
\ee 
attained when $x=1$ and $\alpha=\frac{\pi}{4}$.
Provided that $\eta_0\neq\eta_1$ and $\eta_0\neq\frac{\pi}{2}-\eta_1$, we have the probability of success with feedback greater than without feedback as shown in Fig.\ref{fig6}.

We note that the best measurement on the environment is on the basis $|\pm\rangle\equiv(|0\rangle\pm|1\rangle)/\sqrt{2}$ and the best input state is always $|1\rangle$ in contrast to what happened in the absence of feedback.
In Fig.\ref{fig6} it is also visible a slight asymmetry of the behavior { between
the regions $\eta_1<\pi/2-\eta_0$ and $\eta_1>\pi/2-\eta_0$.
We suspect that this is due to the fact that 
projection on the environment helps in creating states on the main system that are better distinguishable, provided that system and environment are enough entangled prior the measurement. For big $\eta$'s they are not, hence the process is less effective.
}

Finally, it is worth mentioning a similarity of this problem with the discrimination of unitary dilations of the amplitude damping channel using local measurements (analogous similarity was pointed out in Ref.\cite{QIP19} for channel estimation).


\section{Two-shot discrimination}\label{sec:twoshot}

In this section, we shall consider the discrimination assuming to have two copies of the channel characterized either by parameter $\eta_0$ or by parameter $\eta_1$ (with again equal probability $1/2$).

\subsection{Optimal input}\label{subsec:opt2}

Let us first consider the possibility of using entangled inputs.
Since for a single shot the optimal input state lies in the $x-z$ plane, we construct entangled input states as a linear combination of 2-fold tensor product of these states with real coefficients. This amounts to consider the two inputs state as
\begin{equation}\label{vec1}
\sqrt{1-x}|01\rangle+\sqrt{x}|10\rangle,
\end{equation}
or 
\begin{equation}\label{vec2}
\sqrt{1-x}|00\rangle+\sqrt{x}|11\rangle,
\end{equation}
with $x\in[0,1]$.

According to the 2-fold action of \eqref{calN}, in case of \eqref{vec1} we will get the output states as
\begin{align}\label{Bell1}
\rho_i^{(2)}&=\sin^2\eta_i |00\rangle\langle 00|
+(1-x) \cos^2\eta_i  |01\rangle\langle 01| 
+ x \cos^2\eta_i  |10\rangle\langle 10| 
\notag\\
&+\sqrt{x(1-x)} \cos^2\eta_i \left(    |01\rangle\langle 10| +|10\rangle\langle 01| \right),\quad i=0,1.
\end{align}
While in case of \eqref{vec2} as
\begin{align}\label{Bell2}
\rho_i^{(2)}&=\left((1-x)+x\sin^4\eta_i\right)\eta |00\rangle\langle 00|
+x \sin^2\eta_i\cos^2\eta_i  \left( |01\rangle\langle 01| + |10\rangle\langle 10| \right)
+ x \cos^4\eta_i |11\rangle\langle 11|   
\notag\\
&+\sqrt{x(1-x)} \cos^2\eta_i \left(    |00\rangle\langle 11| +|11\rangle\langle 00| \right), \quad i=0,1.
\end{align}
Using \eqref{sucpr} with the states \eqref{Bell1} gives 
\begin{equation}
P_{succ}=\frac{1}{2}\left(1+\sin^2\eta_0-\sin^2\eta_1\right),
\end{equation}
which is independent of $x$.
On the other hand, using \eqref{sucpr} with the states \eqref{Bell2} gives
{ 
\begin{align}
P_{succ}=\frac{1}{2}&\Bigg\{1+ \frac{x}{4} \left|\cos^2(2\eta_0)-\cos^2(2\eta_1)\right| 
+\frac{\sqrt{x}}{2} \left| \cos(2\eta_0)-\cos(2\eta_1)\right| 
\Bigg\},
\end{align}
}
which attains its maximum for $x=1$.
Therefore, we can conclude that entanglement across the two inputs is useless.
Then, we consider as input the 2-fold tensor product of \eqref{ketpsi} 
and optimize over $x$. In other words, we consider
\begin{equation}\label{Psucc2}
P_{succ}=\max_x \frac{1}{2}\left(1+\frac{1}{2} \Big\| \rho_0\otimes\rho_0 -\rho_1\otimes\rho_1
\Big\|_1\right),
\end{equation}
where $\rho_i$s are given by \eqref{rho2}. 

\begin{figure}[H]
\centering
\includegraphics[width=8cm]{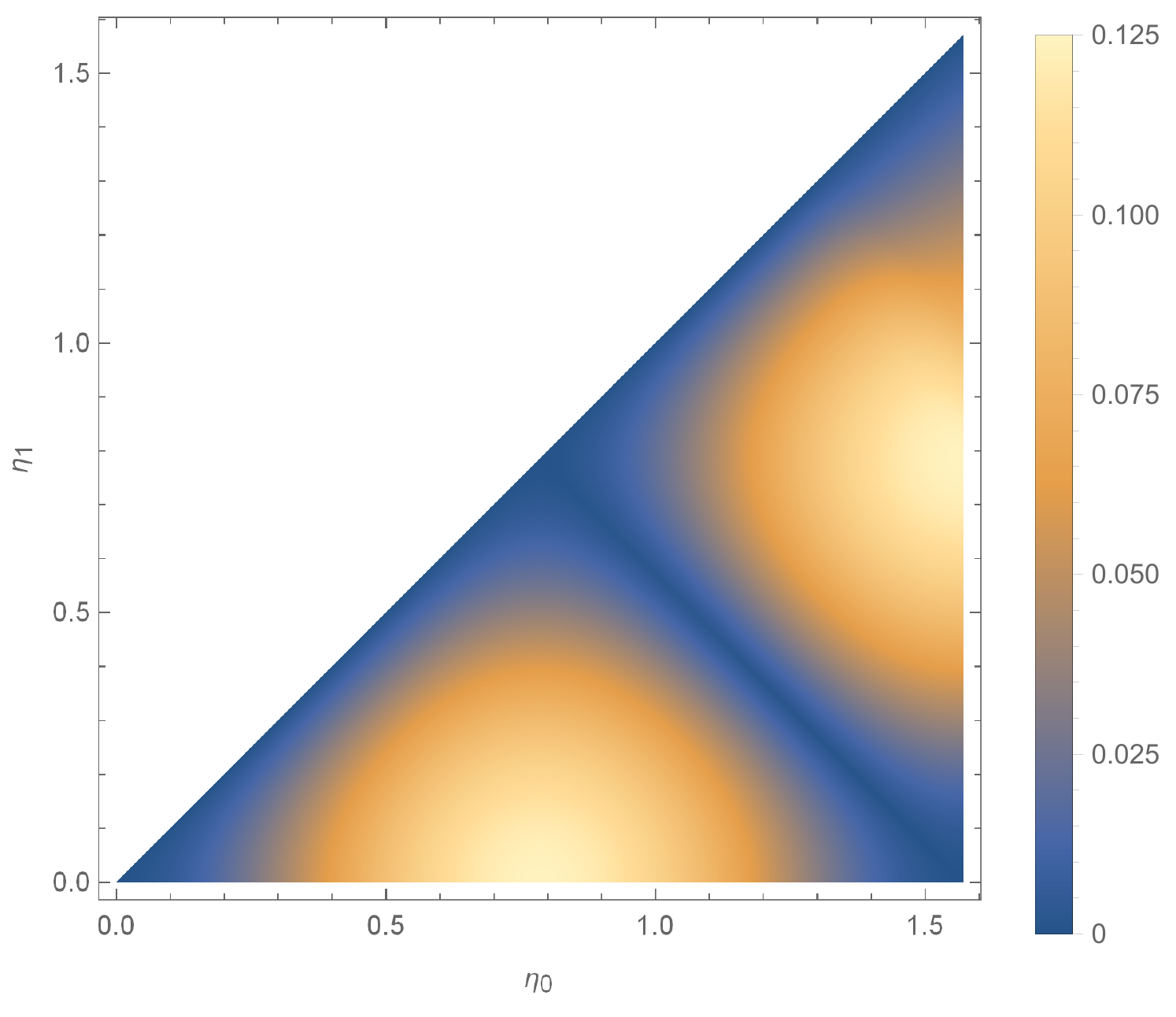}
\caption{Difference between the maximum probability of success \eqref{Psucc2} and \eqref{Psuccnofb} vs $\eta_0$ and $\eta_1$. }
\label{fig7}
\end{figure}

The difference between this optimized success probability and \eqref{Psuccnofb} is shown in Fig.\ref{fig7}.

\begin{figure}[H]
\centering
\includegraphics[width=8cm]{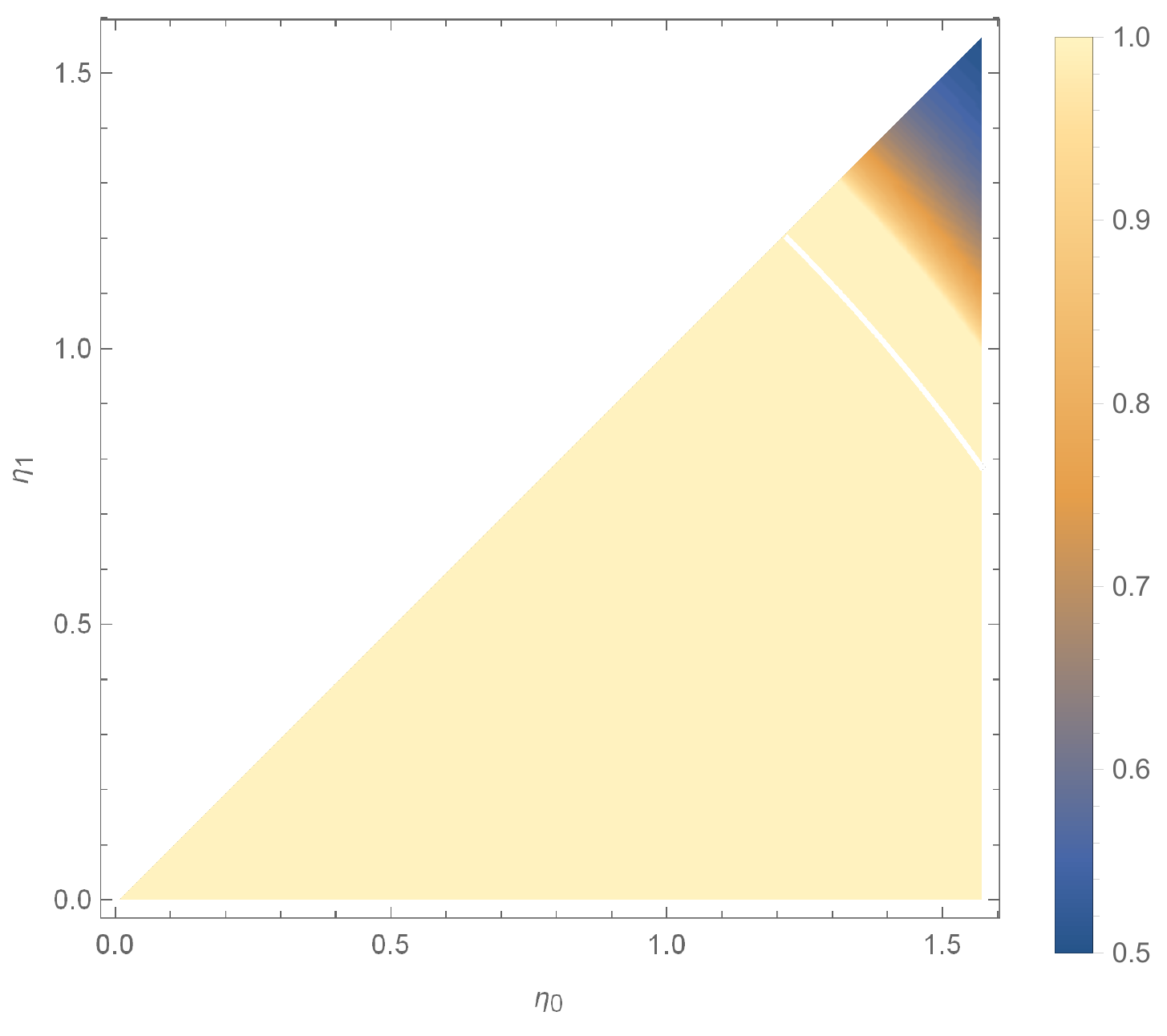}
\caption{Optimal value $x^*$ of $x$ for \eqref{Psucc2} vs  $\eta_0$ and $\eta_1$. The {white} line represents the boundary $\gamma=1/\sqrt{2}$ for the single-shot case.}
\label{fig8}
\end{figure}

The optimal value $x^*$ of $x$ for \eqref{Psucc2} as function of $\eta_0$ and $\eta_1$ is reported in Fig.\ref{fig8}. Note that the region where the $x^*$ value is smaller than one is shrunk with respect to the single-shot case ($\gamma<1/2$). 
Although the exact boundary cannot be expressed analytically, 
{we found numerically that}
\begin{equation}
\gamma(\eta_0,\eta_1)< \frac{1}{2} \Rightarrow x<1,
\end{equation}
where $\gamma(\eta_0,\eta_1)$ is given by \eqref{gamma}. It is worth remarking that for $x^*=1$ the optimal observable $\rho_0-\rho_1$ constructed with \eqref{Bell1} turns out to be local (its normalized eigenvectors are $|00\rangle, |01\rangle, |10\rangle, 
|11\rangle$), while for $x^*<1$ results nonlocal (its normalized eigenvectors are entangled).

\subsection{Adaptive strategy}\label{subsec:ada}

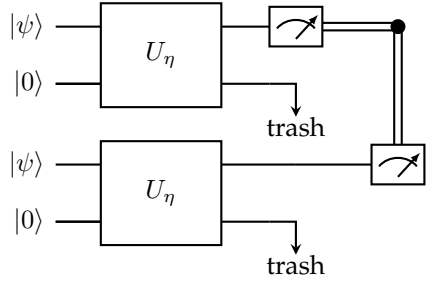
\begin{figure}[ht]
\centering
\begin{quantikz}[row sep=0.01cm,column sep=0.6cm]
\lstick{$\ket{\psi}$}& \gate[wires=2]{U_{\eta}}{0.2cm} & \meter{}&  \cwbend{2} \\ 
\lstick{$\ket{0}$}&\qw& \trash{\text{trash}} \\
\lstick{$\ket{\psi}$}& \gate[wires=2]{U_{\eta}}{0.2cm}& \qw & \meter{}  \\ 
\lstick{$\ket{0}$}&\qw& \trash{\text{trash}} \\
\end{quantikz}    
      \caption{Schematic representation of channel discrimination through a local adaptive strategy. The measurement outcome on the first copy determines the measurement to be performed on the second copy. }
      \label{fig9}
\end{figure}

In the previous subsection, although getting rid of entangled inputs, we saw the necessity of using collective measurement in some parts of the parameters' region.
What happens if we restrict to individual measurements to have a completely local strategy (analogously to \cite{Tej})?
We expect an improvement with respect to the one shot-case, but this relies on using an adaptive strategy, which can generally be depicted as in Fig.\ref{fig9}.

At the output of the two channel copies, we have 
\begin{equation}\label{nmixture}
\frac{1}{2} \rho_0^{\otimes 2} + \frac{1}{2} \rho_1^{\otimes 2},
\end{equation}
with $\rho_i$ given by \eqref{rho2}.

On the first copy we use the POVM whose elements are $|v_0\rangle \langle v_0|$ and
$|v_1\rangle \langle v_1|$, with
$|v_0\rangle$ and $|v_1\rangle$ the (normalized) 
eigenvectors of the observable $\rho_0-\rho_1$.
On the second copy, we choose a POVM  depending on the previous copy's measurement outcome, i.e., $\Pi_{x_2}^{x_1}$, $x_i\in\{0,1\}$, where the subscript denotes the element of the POVM, while the superscript the dependence from the previous measurement outcome.   
We define
\be
P_j(x_2|x_{1})\equiv{\rm Tr}\left[\rho_j \Pi_{x_2}^{x_{1}}\right].
\ee
Then the success probability will be given by
\begin{align}\label{Ptomax}
P_{succ}&=\frac{1}{2}\left[P_0(0|0)\langle v_0 |\rho_0 | v_0\rangle 
+ P_0(0|1) \left(1-\langle v_0 |\rho_0 | v_0\rangle\right)\right] \notag\\
&+\frac{1}{2}\left[P_1(1|1)\langle v_1 |\rho_1 | v_1\rangle 
+ P_1(1|0) \left(1-\langle v_1 |\rho_1 | v_1\rangle\right)\right]. 
\end{align}
Eq.\eqref{Ptomax} should be maximized overall POVMs $\Pi_{x_2}^{x_1}$.
Actually, the first and fourth terms can be maximized overall
$\Pi_0^0$ being $\Pi_1^0=I-\Pi_0^0$,
while the second and third terms can be maximized overall 
$\Pi_1^1$ being $\Pi_0^1=I-\Pi_1^1$.

{
So we can independently perform the following maximizations (for fixed input $x$): 
\begin{align}\label{Pi00}
\max_{\Pi_0^0 : 0\leq \Pi_0^0 \leq I} \frac{1}{2}\left\{
{\rm Tr}\left[\rho_0 \Pi_0^0\right] \langle v_0 |\rho_0 | v_0\rangle 
+ {\rm Tr}\left[\rho_1(I-\Pi_0^0)\right] \left(1-\langle v_1 |\rho_1 | v_1\rangle\right)\right\},
\end{align}
and
\begin{align}\label{Pi11}
\max_{\Pi_1^1 : 0\leq \Pi_1^1 \leq I} \frac{1}{2}\left\{
 {\rm Tr}\left[\rho_0(I-\Pi_1^1)\right] \left(1-\langle v_0 |\rho_0 | v_0\rangle\right)
+{\rm Tr}\left[\rho_1 \Pi_1^1\right] \langle v_1 |\rho_1 | v_1\rangle 
\right\}.
\end{align}
Let us consider the problem of Eq.\eqref{Pi00}.
Defining $p_0:=\langle v_0 |\rho_0 | v_0\rangle$ and $q_0:=\langle v_0 |\rho_1 | v_0\rangle$ 
we can recast it into the following form
\begin{align}
\frac{p_0+q_0}{2}
\max_{\Pi_0^0 : 0\leq \Pi_0^0 \leq I} 
\left\{
\frac{p_0}{p_0+q_0} {\rm Tr}\left[\rho_0 \Pi_0^0\right] 
+ \frac{q_0}{p_0+q_0}{\rm Tr}\left[\rho_1(I-\Pi_0^0)\right] \right\},
\end{align}
which coincides with the optimal discrimination of $\rho_0$ and $\rho_1$, this time appearing with probabilities $\frac{p_0}{p_0+q_0}$ and $\frac{q_0}{p_0+q_0}$ respectively. Hence, it can be solved by again resorting to Helstrom measurement. In other words, $\Pi_0^0$ and $I-\Pi_0^0$ can be constructed as projectors onto positive and negative subspaces of the operator
$\frac{p_0}{p_0+q_0} \rho_0 - \frac{q_0}{p_0+q_0}\rho_1$.

With Eq.\eqref{Pi11} one can proceed similarly by defining 
$p_1:=\langle v_1 |\rho_1 | v_1\rangle$ and $q_1:=\langle v_1 |\rho_0 | v_1\rangle$. 
Therefore, $\Pi_1^1$ and $I-\Pi_1^1$ will be constructed as projectors onto positive and negative subspaces of the operator
$\frac{q_1}{p_1+q_1} \rho_0 - \frac{p_1}{p_1+q_1}\rho_1$.
Ultimately, we can get the success probability as the function 
$P_{succ}(\eta_0,\eta_1,x)$ given by
\begin{align}\label{PsMold}
P_{succ}&=\frac{p_0+q_0}{2} \frac{1}{2}\left\{1+\left\|
\frac{p_0}{p_0+q_0} \rho_0 - \frac{q_0}{p_0+q_0}\rho_1 \right\|_1\right\}\notag\\
&+\frac{p_1+q_1}{2} \frac{1}{2}\left\{1+\left\|
\frac{q_1}{p_1+q_1} \rho_0 - \frac{p_1}{p_1+q_1}\rho_1 \right\|_1\right\}.
\end{align}
}

\begin{figure}[H]
\centering
\includegraphics[width=8cm]{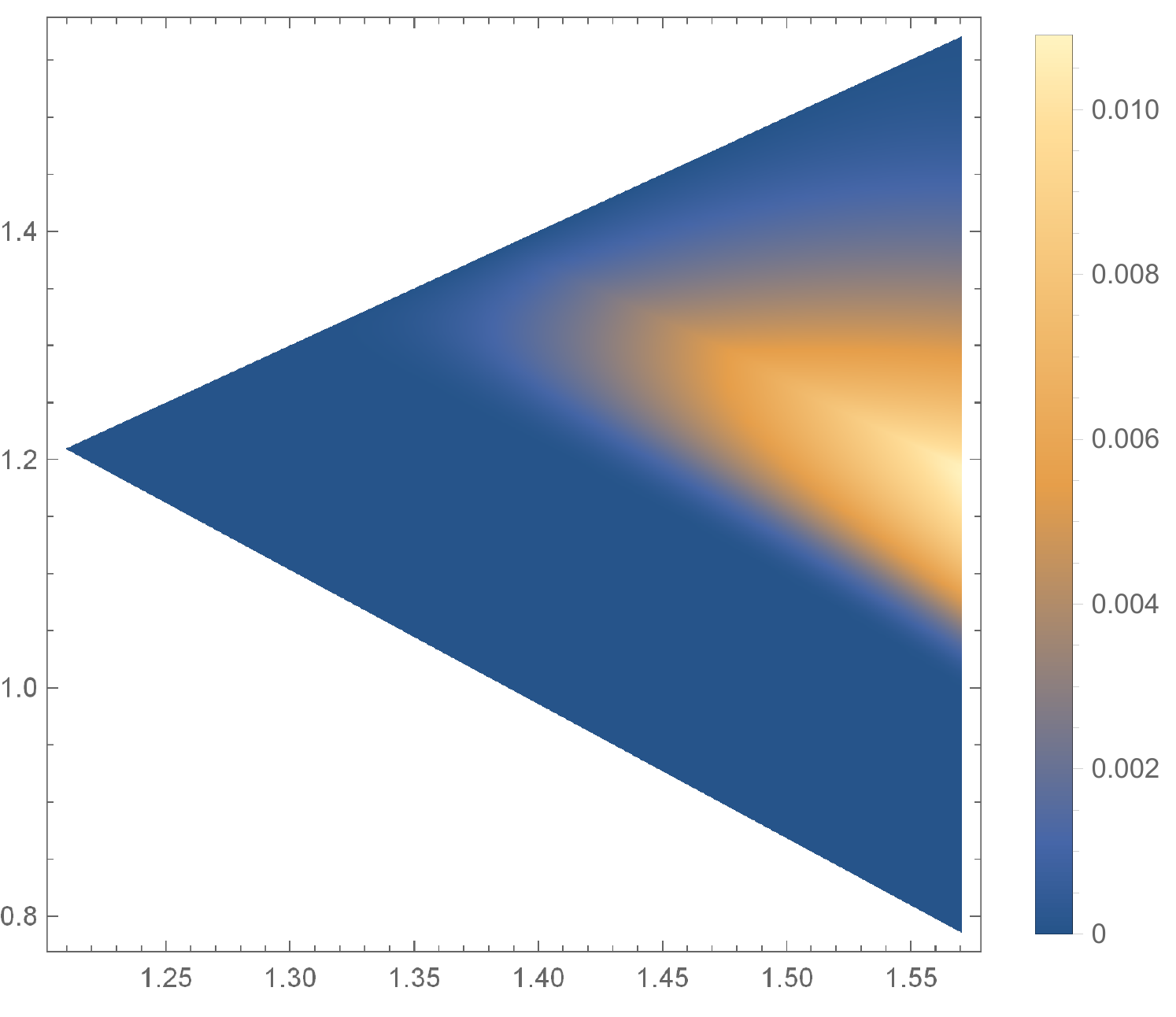}
\caption{Difference between the maximum probability of success using collective measurement  (Eq.\eqref{Psucc2}) and that  using the adaptive strategy (Eq.\eqref{PsMold} maximized over $x$) 
vs $\eta_0$ and $\eta_1$ in the region $\gamma<1/\sqrt{2}$. }
\label{fig10}
\end{figure}

To evaluate the performance of the adaptive strategy, we compare this probability, maximized over $x$, 
with the optimal success probability for two-shot \eqref{Psucc2}. 
Fig.\ref{fig10} shows the difference between the latter and the former. 
As we expected, such a difference is nonzero only in the region of Fig.\ref{fig8} where $x^*<1$, 
however it is very tiny. This shows that the devised local adaptive strategy performs almost like the 
strategy involving collective measurement. 

\begin{figure}[H]
\centering
\includegraphics[width=8cm]{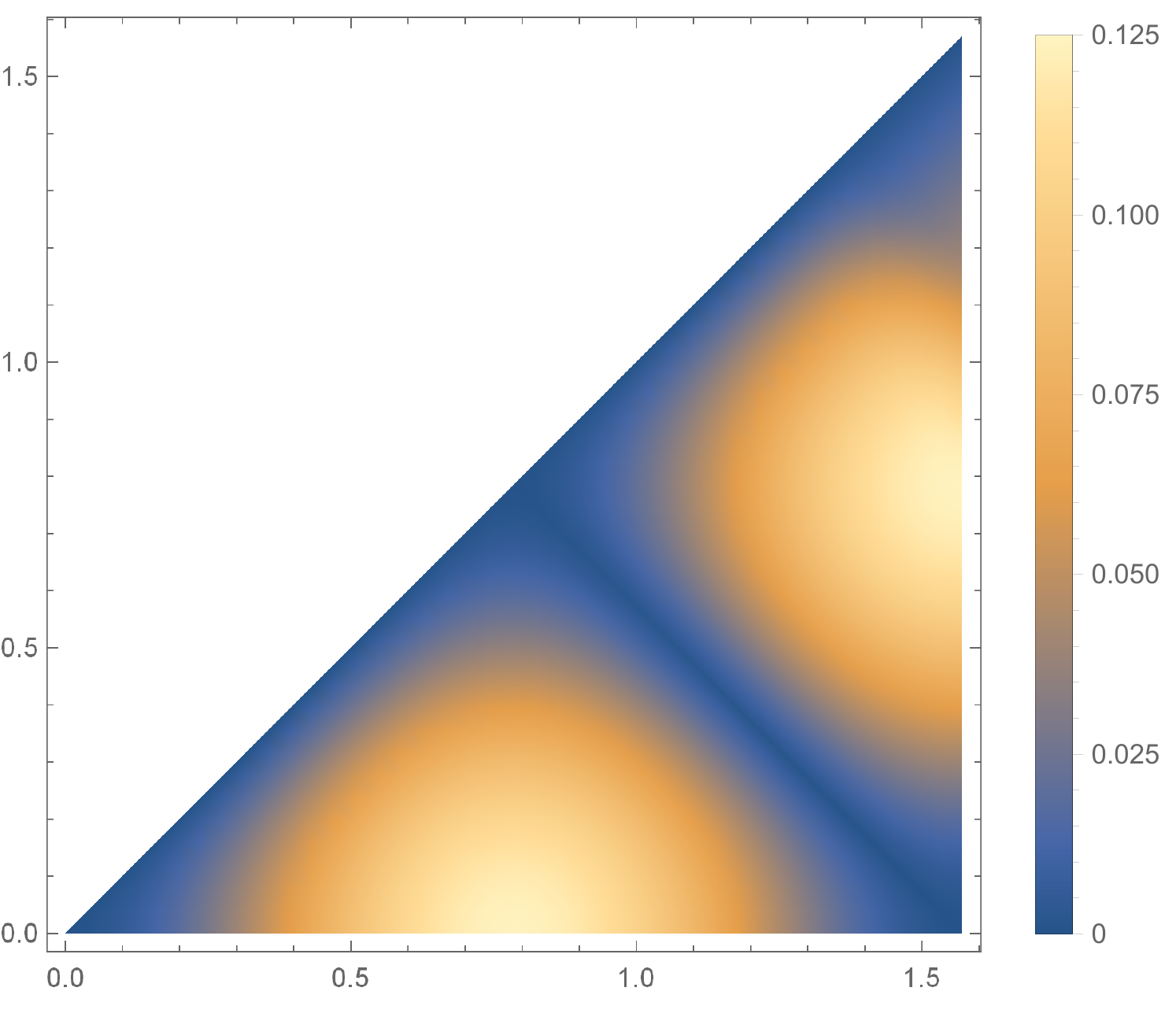}
\caption{
Difference between the maximum probability of success using the adaptive strategy  (Eq.\eqref{PsMold} maximized over $x$) 
and that of a single shot (Eq.\eqref{Psuccnofb}) vs $\eta_0$ and $\eta_1$. }
\label{fig11}
\end{figure}

\noindent Furthermore, Fig.\ref{fig11} shows the difference between the maximum probability of success using the adaptive strategy and that of a single-shot. Note that the range of values in Fig.\ref{fig11} is one order of magnitude bigger than Fig.\ref{fig10}. {As a consequence Fig.\ref{fig11} faithfully follows Fig.\ref{fig7}. }

{ 

We close this SubSection saying that the strategy we pursued was based on 
Bayesian updating using forward optimization, much like Ref.\cite{acin05}. 
However in Ref.\cite{hig11} it was shown that backward optimization can 
give better performance in case mixed states have to be discriminated.
In the present case the latter seems not much effective. Anyway,
a comparison of the two methods is provided in Appendix \ref{optPOVM}.
}


\subsection{Adaptive strategy with feedback}\label{subsec:adafb}

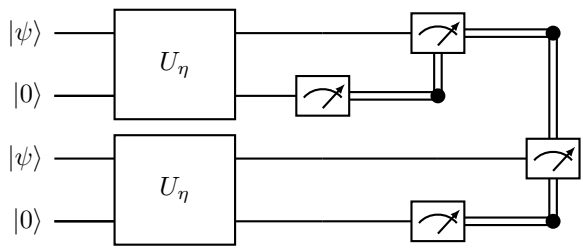
\begin{figure}[ht]
\centering
\begin{quantikz}[row sep=0.16cm,column sep=0.8cm]%
\lstick{$\ket{\psi}$}& \gate[wires=2]{U_{\eta}}{0.2cm} &\qw& \meter{}&  \cwbend{2} \\ 
\lstick{$\ket{0}$}&\qw& \meter{}& \cwbend{-1} &&\\
\lstick{$\ket{\psi}$}& \gate[wires=2]{U_{\eta}}{0.2cm}& \qw &\qw& \meter{}  \\ 
\lstick{$\ket{0}$}&\qw&\qw& \meter{} & \cwbend{-1}\\
\end{quantikz}    
      \caption{Schematic representation of channel discrimination through a local adaptive strategy including feedback. On each step the measurement to be performed on the main system is determined by 
the measurement outcome of the environment on that step, together with       
the measurement outcome of the main system at the previous step.}
      \label{fig12}
\end{figure}

We now develop an adaptive strategy that includes the feedback from the environment,
as illustrated in Fig.\ref{fig12}.

On the first copy of the unitary, we proceed like in Subsection \ref{subsec:fb}, 
defining a POVM $\Pi_{x_1}^{e_1}$ whose elements are
\begin{align}
\Pi_{x_1=0}^{e_1}&=|v_0^{e_1}\rangle
\langle v_0^{e_1} |, \\
\Pi_{x_1=1}^{e_1}&=|v_1^{e_1}\rangle
\langle v_1^{e_1} |, 
\end{align}
where $|v_0^{e_1}\rangle$ 
and $|v_1^{e_1}\rangle$ 
are (normalized) eigenvectors of $\varphi_0^{e_1}-\varphi_1^{e_1}$.
Here $\varphi_j^{e_1}\equiv |\varphi_j^{e_1}\rangle\langle\varphi_j^{e_1}|$ with $j=0,1$ 
and $e_1=\pm$. Note that following the conclusions of Subsection \ref{subsec:fb},
we are considering the input $|\psi\rangle$ as $|1\rangle$ and the measurement on the environment on the basis $|\pm\rangle$, so that
\begin{subequations}
\begin{align}
|\varphi_0^\pm\rangle&=\mp i\sin\eta_0 |0\rangle+\cos\eta_0 |1\rangle, \\
|\varphi_1^\pm\rangle&=\mp i\sin\eta_1 |0\rangle+\cos\eta_1 |1\rangle. 
\end{align}
\end{subequations}
As a consequence, at the output of the main system, we have to distinguish between pure states (differently to what happened in the previous Subsection). 
{ 
We then define the probability
\be\label{condprob}
P^{e_2}_j(x_2|x_1,e_2)\equiv
{\rm Tr}\left[\varphi_j^{e_2} \, \Pi_{x_2}^{x_1,\, e_2}\right],
\ee
where $\Pi_{x_2}^{x_1,\, e_2}$ indicates the POVM elements characterized by
$x_i\in\{0,1\}$ and $e_2=\pm$.
}

\begin{figure}[H]
\centering
\includegraphics[width=8cm]{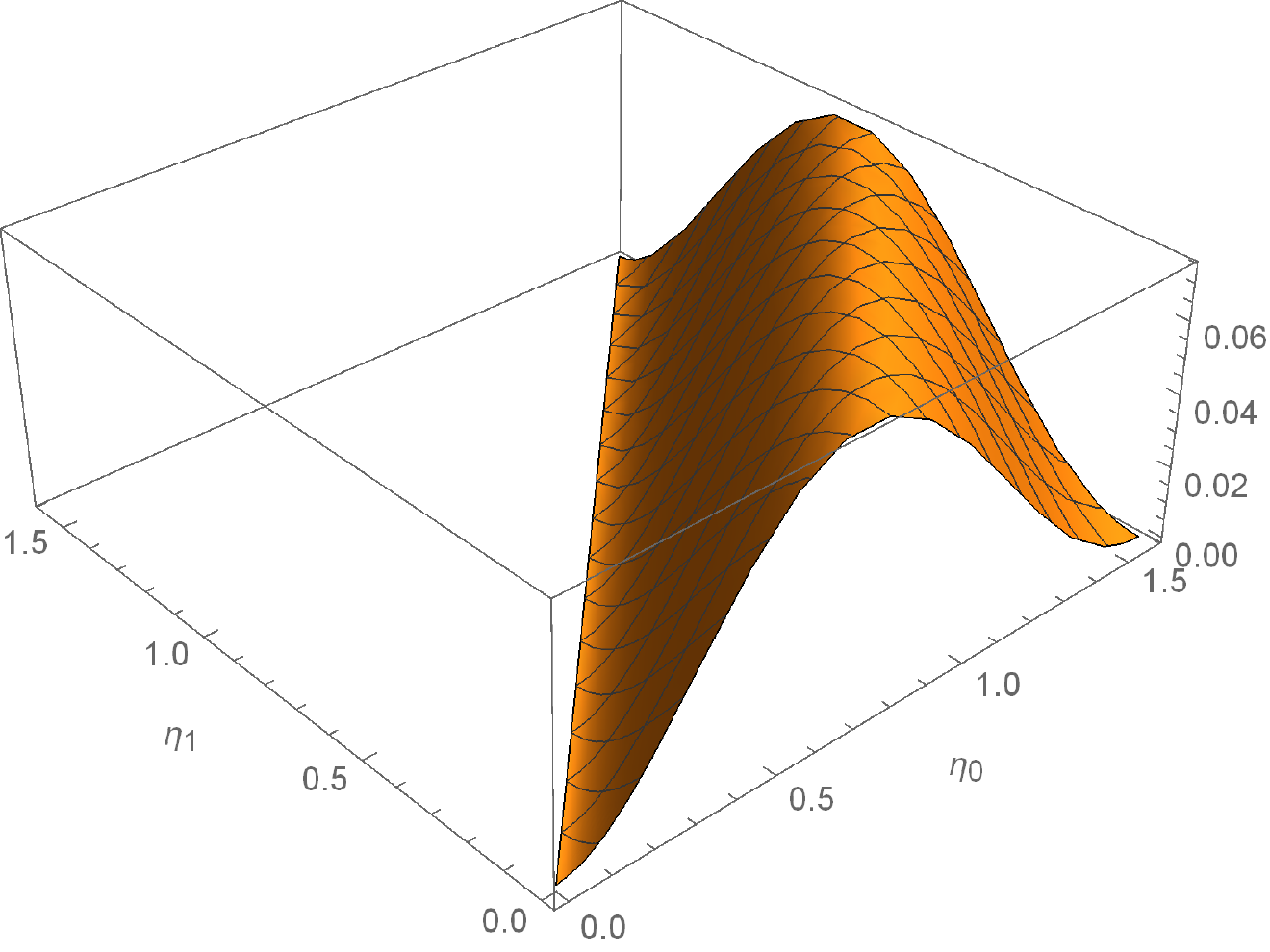}
\caption{Difference between the probabilities of success \eqref{Psuccadafb} and \eqref{Psuccfb} vs $\eta_0$ and $\eta_1$.}
\label{fig13}
\end{figure}

In terms of it, we can express the success probability as
\begin{align}
P_{succ}=\sum_{e_1,e_2=\pm}\frac{1}{8}\Big[&
P_0^{e_2}(0|0,e_2)\langle v_0^{e_1} |\varphi_0^{e_1} | v_0^{e_1}\rangle
+P_0^{e_2}(0|1,e_2)\left(1-\langle v_0^{e_1} |\varphi_0^{e_1} | v_0^{e_1}\rangle\right) \notag\\
&+P_1^{e_2}(1|1,e_2)\langle v_1^{e_1} |\varphi_1^{e_1} | v_1^{e_1}\rangle
+P_1^{e_2}(1|0,e_2)\left(1-\langle v_1^{e_1} |\varphi_1^{e_1} | v_1^{e_1}\rangle\right)\Big],
\end{align}
where the factor $\frac{1}{8}$ in front of the square brackets arises from the probability $\frac{1}{2}$ of having $U_{\eta_i}\otimes U_{\eta_i}$, the probability $\frac{1}{2}$ for $e_1$ to take one of the two values, and the probability $\frac{1}{2}$ for $e_2$ to take one of the two values.
{
Now for each value of $e_1$ and $e_2$ we can separately maximize the terms
\be
\max_{\Pi_0^{0,\, e_2} : 0\leq \Pi_0^{0,\, e_2} \leq I} \frac{1}{8}\left\{
{\rm Tr}\left[\varphi_0^{e_2}  \Pi_{0}^{0,\, e_2}\right] 
\langle v_0^{e_1} |\varphi_0^{e_1} | v_0^{e_1}\rangle
+{\rm Tr}\left[\varphi_0^{e_2} \left(I- \Pi_{0}^{0,\, e_2}\right)\right] 
\left(1-\langle v_1^{e_1} |\varphi_1^{e_1} | v_1^{e_1}\rangle\right) \right\},
\ee
and
\be
\max_{\Pi_1^{1,\, e_2} : 0\leq \Pi_1^{1,\, e_2} \leq I} \frac{1}{8}\left\{
{\rm Tr}\left[\varphi_0^{e_2}  \left(I-\Pi_{1}^{1,\, e_2}\right)\right] 
\langle v_0^{e_1} |\varphi_0^{e_1} | v_0^{e_1}\rangle
+{\rm Tr}\left[\varphi_0^{e_2} \Pi_{1}^{1,\, e_2}\right] 
\left(1-\langle v_1^{e_1} |\varphi_1^{e_1} | v_1^{e_1}\rangle\right) \right\},
\ee
following the same reasoning of SubSec.\ref{subsec:ada}.

Ultimately, we arrive at 
\begin{align}\label{Psuccadafb}
P_{succ}=\frac{1}{2}\left[1+\sin\left(\eta_0-\eta_1\right)\sqrt{1+\cos^2(\eta_0-\eta_1)}\right].
\end{align}
}
The improvement with respect to the single-shot with feedback (Eq.\eqref{Psuccfb}) is shown in Fig.\ref{fig13}. We can see that it is positive in all parameters' region (but $\eta_1=\eta_0$ and 
$\eta_0=\frac{\pi}{2},\eta_1=0$) and is maximum for $\eta_1=\eta_0-\frac{\pi}{4}$.

\section{Concluding Remarks}\label{sec:con}

We have addressed the issue of discriminating between two-qubit amplitude damping channels by considering single and double-shot.
For the one-shot, we showed that the excited state as input is not always optimal, and the side entanglement assistance has a limited benefit.
In contrast, feedback assistance from the environment is highly beneficial.
About feedback, we suspect that it will be even more effective in case of discrimination of random unitary channels, where the information recovered from the environment neutralizes the channel action (taking it back to identity map).

For the two-shot, we proved the in-utility of entangled inputs. 
Then, focusing on individual (local) measurements, we found the optimal adaptive strategy.
We are confident that this strategy can be extended in a Markovian way to $n$-shot, and the asymptotic analysis of its performance would be foreseeable, {eventually by employing techniques like dynamical programming or reinforcement learning \cite{brand19,bilkis20}.}
Similarly, the adaptive strategy is applied to environment feedback assisted discrimination showing a smaller improvement than the case without feedback (cfr. Figs.\ref{fig11} and \ref{fig13}). This is because feedback has provided a big enhancement already on the first shot.

{
It is worth remarking that having two copies of the channel one could have used them in sequence, not in parallel as done in Sec.\ref{sec:twoshot}. In such a case one is led to distinguish between ${\cal N}(\rho_0)$ and ${\cal N}(\rho_1)$ (where $\rho_i$s are given in \eqref{rho2}), each occurring with probability $\frac{1}{2}$. Simple calculations show that this latter strategy cannot be better than the parallel one.
For example, without feedback, its success probability equals \eqref{PsMold} in the region 
$\eta_1<\eta_0$ and $\eta_1<\frac{\pi}{2}-\eta_0$, while it results strictly smaller than  \eqref{PsMold} in the region $\eta_1<\eta_0$ and $\eta_1>\frac{\pi}{2}-\eta_0$.
}

In the future, it is worth extending the performed analysis to $d$-dimensional amplitude damping channels, also with restrictions on the set of input states (e.g., energy restriction).
After all, investigating amplitude damping channel discrimination in discrete systems can also provide new insights for distinguishing continuous variable lossy channels \cite{Inv},
{although this is often traced back to the distinguishability of coherent states (for this latter subject see e.g. \cite{ass11,dolinar,bilkis20}).}



\appendix

\section{Optimal POVMs for local adaptive strategy}\label{optPOVM}

\begin{figure}[H]
\centering
\includegraphics[width=8cm]{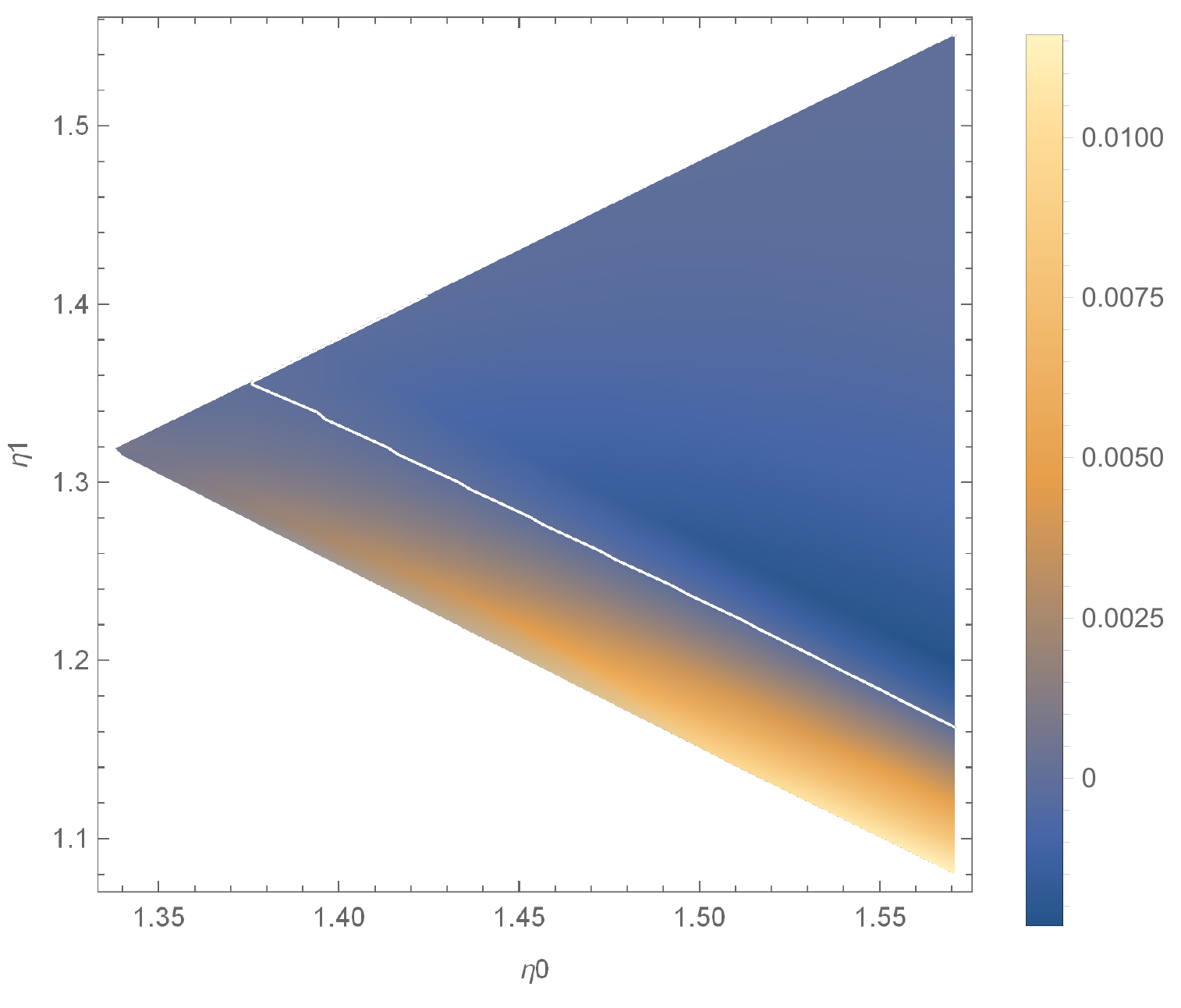}
\caption{Difference \eqref{fwdbwd} between forward and backward optimized success 
probabilities vs $\eta_0$ and $\eta_1$. The white line represents 0. Only the region where 
$P^{(\rightarrow)}_{succ}(\eta_0,\eta_1,x)$ differs form \eqref{Psucc2} is shown.}
\label{fig15}
\end{figure}

{ 
Suppose that instead of projecting onto $|v_0\rangle$ and $|v_1\rangle$ on the first copy we realize a measurement by a POVM $\{M,I-M\}$. Then 
\eqref{Ptomax} will become
\begin{align}\label{PsM}
P_{succ}=\frac{1}{2}&\left\{
{\rm Tr}\left[ \rho_0 \Pi_0^0\right] {\rm Tr}\left[ \rho_0 M\right]
+{\rm Tr}\left[ \rho_0 \left(I-\Pi_1^1\right)\right] {\rm Tr}\left[ \rho_0 \left(I-M\right)\right]
\right.\notag\\
&\left. 
+{\rm Tr}\left[ \rho_1 \Pi_1^1 \right] {\rm Tr}\left[ \rho_1 \left(I-M\right)\right]
+{\rm Tr}\left[ \rho_1 \left(I-\Pi_0^0\right)\right] {\rm Tr}\left[ \rho_1 M\right] 
\right\}.   
\end{align}
Defining
\begin{align}
&r_0\equiv {\rm Tr}\left[ \rho_0 M\right], \qquad s_0\equiv {\rm Tr}\left[ \rho_1 M\right],\\
&r_1\equiv 1-s_0, \qquad \quad s_1=1-r_0,
\end{align}
equation \eqref{PsM} can be rewritten as 
\begin{align}\label{PsMnew}
P_{succ}&=\frac{r_0+s_0}{2}\left\{ \frac{r_0}{r_0+s_0}
{\rm Tr}\left[ \rho_0 \Pi_0^0\right] 
+\frac{s_0}{r_0+s_0}{\rm Tr}\left[ \rho_1 \left(I-\Pi_0^0\right)\right] \right\}\notag\\
&+\frac{r_1+s_1}{2}\left\{ 
\frac{s_1}{r_1+s_1}{\rm Tr}\left[ \rho_0 \left(I-\Pi_1^1\right)\right] 
+\frac{r_1}{r_1+s_1}{\rm Tr}\left[ \rho_1 \Pi_1^1 \right]
\right\}.   
\end{align}
Let us now impose that the measurement on the second copy is of the Helstrom kind.
Thus $\Pi_0^0$ and $(I-\Pi_0^0)$ are projectors onto the positive and negative eigenspaces of 
\be
 \frac{r_0}{r_0+s_0}\rho_0-  \frac{s_0}{r_0+s_0}\rho_1.
\ee
Analogously 
$\Pi_1^1$ and $(I-\Pi_1^1)$ are projectors onto the positive and negative eigenspaces of 
\be
 \frac{r_1}{r_1+s_1}\rho_0-  \frac{s_1}{r_1+s_1}\rho_1.
\ee
As such all the quantities appearing in Eq.\eqref{PsMnew} will depend upon $M$ (for a fixed $x$).
Then, maximizing $P_{succ}$ over $M$, such that $0\leq M\leq I$, realizes the backward optimization.

In such a way we can get the success probability $P^{(\leftarrow)}_{succ}(\eta_0,\eta_1,x)$.
We also denote by $P^{(\rightarrow)}_{succ}(\eta_0,\eta_1,x)$ the quantity \eqref{PsMold}.
Then, in Fig.\ref{fig15}, we show the difference
\begin{align}\label{fwdbwd}
\max_x P^{(\rightarrow)}_{succ}(\eta_0,\eta_1,x)-\max_x P^{(\leftarrow)}_{succ}(\eta_0,\eta_1,x).
\end{align}
We may note that when the backward optimization performs better, it only gives 
a tiny improvement (of the order of $10^{-3}$) to the success probability. 
}

\bigskip
\bigskip

The authors equally contributed to this work.
\\

{
We acknowledge the funding from the European Union’s
Horizon 2020 research and innovation programme under grant agreement
No 862644 (FET-Open project: QUARTET).}
\\

{The authors thank Francisco Revson F. Pereira for useful discussions.}



\end{document}